\documentclass[sigconf]{acmart}
\renewcommand\footnotetextcopyrightpermission[1]{} 
\usepackage{makecell}
\usepackage{multirow}
\usepackage{soul}
\usepackage{CJKutf8}
\usepackage{graphicx}
\usepackage{subcaption}

\AtBeginDocument{%
  }

\begin{document}

\title{QExplorer: Large Language Model Based Query Extraction for Toxic Content Exploration}

\author{Shaola Ren}
\affiliation{%
  \institution{Alibaba Group}
  \city{Hangzhou}
  \state{Zhejiang}
  \country{China}
}
\email{shaola.rs@alibaba-inc.com}

\author{Li Ke}
\affiliation{%
  \institution{Alibaba Group}
  \city{Hangzhou}
  \state{Zhejiang}
  \country{China}
}
\email{keli.kl@alibaba-inc.com}

\author{Longtao Huang}
\affiliation{%
  \institution{Alibaba Group}
  \city{Hangzhou}
  \state{Zhejiang}
  \country{China}
}
\email{kaiyang.hlt@alibaba-inc.com}

\author{Dehong Gao}
\affiliation{%
  \institution{Northwestern Polytechnical University}
  \city{Xi’an}
  \state{Shaanxi}
  \country{China}
}
\email{dehong.gdh@nwpu.edu.cn}

\author{Hui Xue}
\affiliation{%
  \institution{Alibaba Group}
  \city{Hangzhou}
  \state{Zhejiang}
  \country{China}
}
\email{hui.xueh@alibaba-inc.com}

\renewcommand{\shortauthors}{S Ren et al.}

\begin{abstract}
  Automatically extracting effective queries is challenging in information retrieval, especially in toxic content exploration, as such content is likely to be disguised. With the recent achievements in generative Large Language Model (LLM), we are able to leverage the capabilities of LLMs to extract effective queries for similar content exploration directly. This study proposes \textbf{QExplorer}, an approach of large language model based \textbf{Q}uery Extraction for toxic content \textbf{Explor}ation. The QExplorer approach involves a 2-stage training process: instruction Supervised FineTuning (SFT) and preference alignment using Direct Preference Optimization (DPO), as well as the datasets construction with feedback of search system. To verify the effectiveness of QExplorer, a series of offline and online experiments are conducted on our real-world system. The offline empirical results demonstrate that the performance of our automatic query extraction outperforms that of several LLMs and humans. The online deployment shows a significant increase in the detection of toxic items.
\end{abstract}

\ccsdesc[500]{Information systems~Information retrieval}
\ccsdesc[500]{Computing methodologies~Machine learning}
\ccsdesc[500]{Computing methodologies~Artificial intelligence}

\keywords{query extraction, large language models, supervised fine-tuning}


\maketitle

\section{Introduction}
The content on widely used social media or e-commerce platforms, such as Twitter, Amazon and Taobao, is created by millions of users. It is inevitable that a small portion of toxic content such as offensive, sexual, violent, and fraudulent language will appear. Therefore, the detection and exploration of such toxic content become important for maintaining a healthy platform. This toxic content can appear in the form of text, image, audio, or video. This study focuses on the item's textual information of Xianyu which is the largest Chinese second-hand trading platform.

\begin{figure}[htbp]
  \centering
  \includegraphics[width=\linewidth]{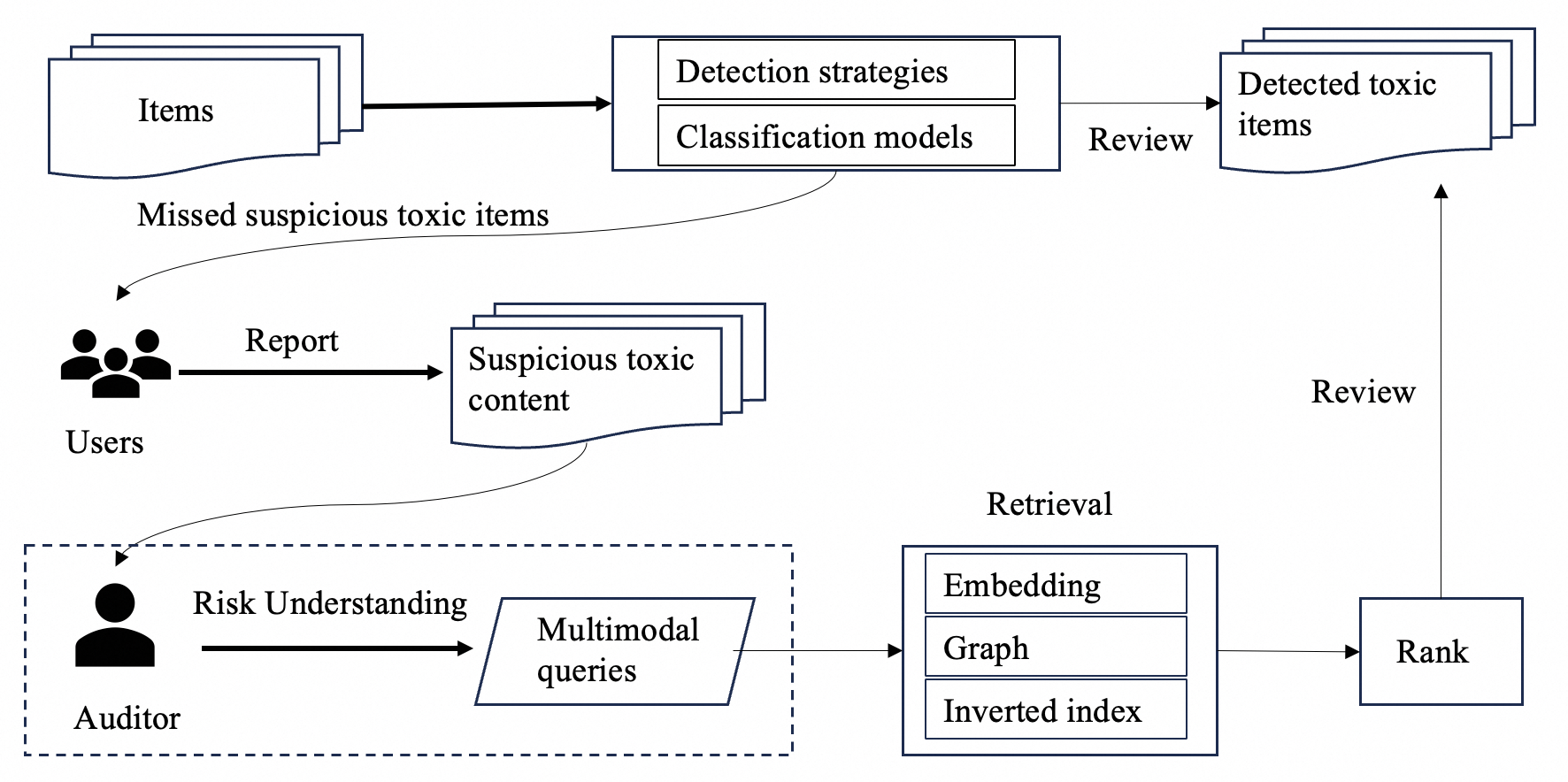}
  \caption{Framework of the toxic content detection system. This system is composed by strategies, classification models, a risk understanding module and a search system. Some suspicious toxic items missed by the strategies and classification models are reported by users of the trading platform. Then the auditors analyze the content and perform multi-modal query searches. The dashed box indicates the auditor risk understanding module, which is the focus of this study.}
  \Description{Framework of the toxic content detection system.}
  \label{fig:toxic_dec_system}
\end{figure}

To recognize toxic content, a general solution is to train classification models to process all the information that is concerned. However, these models cannot identify all toxic items. Figure \ref{fig:toxic_dec_system} illustrates the framework of our toxic content detection system, which contains strategies, classification models, a risk understanding module, and a search system. In our case, some toxic items overlooked by these models are reported by the users of the trading platform. We aim to explore all the similar toxic content when such reports occur. Since toxic content is often camouflaged within normal language expressions, we incorporate a risk understanding module before the search process, as shown in Figure \ref{fig:toxic_dec_system}. The risk understanding module essentially generates queries based on the reported toxic content. Prior to the implementation detailed in this work, only manually annotated queries were used for exploration.

When a suspicious toxic content is reported, the auditor first analyzes the content and extracts keywords or images that need to be explored. The queries are then used to search within a search system comprised of two phases: retrieval and ranking. During the retrieval phase, methods such as inverted index based exact keyword matching, embedding based semantic retrieval, and graph based retrieval are employed. Generally speaking, the inverted index based keyword exact matching is the most effective method for toxic content exploration. In the ranking phase, all retrieved items are ranked according to their risk probabilities, which are calculated by the ranking service.

In this toxic content exploration system shown in Figure \ref{fig:toxic_dec_system}, the auditor risk understanding component serves as the foundation of the entire process. An effective query can result in an efficient toxic content exploration by identifying most of the similar toxic items in the search system. Our auditors are asked to mark out the part related to risk. However, an effective query does not always focus solely on hazardous parts of the content; sometimes specific language patterns are also helpful for exploration, as different expressions with similar meanings are often employed. Table \ref{tab:typical_toxic} shows some typical examples, with italicized or highlighted in red parts indicating the human annotated phrases that are deemed improper or suspicious according to the auditors' judgment. Those annotated phrases are then used for exploring similar toxic content. As observed, not all the annotated phrases are suitable for an efficient search. For example, the phrase ``Transferable Commercial Performance License (2+ Years) and Radio/TV Program Production License'' is overly verbose and may not yield any results in a search system. Meanwhile, the term ``companionship'' is too broad and may result in too many irrelevant items.

\begin{table}
  \caption{Typical toxic content is presented, with the improper segments annotated by humans in the given context. These segments are italicized or highlighted in red.}
  \label{tab:typical_toxic}
  \begin{CJK}{UTF8}{gbsn}
  \begin{tabular}{p{\linewidth}}
    \toprule
    国内外QS排名\textcolor{red}{毕业征}来图\textcolor{red}{复刻}打印定制
    (Customized \textcolor{red}{\textit{Reproduction Printing of Diplomas}} for Domestic or International Universities Ranked by QS) \\
    \midrule
    \textcolor{red}{营业性演出许可(满2年)}+\textcolor{red}{广播节目电视制作(满2年)} 可以转让可以迁入所有城市
    (\textcolor{red}{\textit{Transferable Commercial Performance License (2+ Years) and Radio/TV Program Production License}} (2+ Years), Eligible for Relocation to Any City) \\
    \midrule
    我约您］坐标上海/杭州，本人男单主，周末节假日，约JK或常服。主要活动吃饭逛街看电影等，希望能主打一个\textcolor{red}{陪伴}及情侣感。全程接送一切费用免单。具体收费你开价，只要不太过分都能接受[吉他] 
    (Personal Companion Offer: Male host in Shanghai/Hangzhou for weekends and holidays. Seek company for JK or regular wear. Activities include dining, shopping, and movies with a focus on \textcolor{red}{\textit{companionship}} and a romantic relationship. Round-trip transport and all expenses covered. Rates are negotiable, as long as the demands are not too excessive. [Guitar]) \\
  \bottomrule
\end{tabular}
\end{CJK}
\end{table}

Assistance from models can alleviate the psychological burden on auditors who have been exposed to negative content for a long period. To reduce human dependence in the risk understanding phase and enhance the efficiency in toxic content exploration, we introduce the following LLM based automatic query extraction approach.

Generally, existing methods for keyphrase extracting can be divided into three categories: (1) 2-step keyphrase extraction (\cite{hasan-ng-2014-automatic,7805062,8852151,choi2023simckpsimplecontrastivelearning}) which involves generating a set of phrase candidates using various methods and subsequently ranking or classifying these candidates. (2) Sequence labeling (\cite{10.1145/3308558.3313642,luan-etal-2017-scientific,zhang-etal-2016-keyphrase,Gu_2021}) which predicts the likelihood that a token should be included in the extracted keyphrase. (3) Sequence-to-sequence generation (\cite{wu2024pretrainedlanguagemodelskeyphrase,chowdhury2022applyinggenericsequencetosequencemodel,9443960}) which is based on pretrained models such as Bert, Bart or T5.

The way humans work is fundamentally different from the first two methods. An experienced content safety expert typically has an intuitive sense of which parts of toxic content is efficient for further exploration. This is the intuition of our proposed method.

Recent achievements in LLMs demonstrate their capability to comprehend natural language text and address even more complex real world scenarios. Proprietary LLMs like GPT3.5 or GPT4 have shown impressive performance in query extraction when assisted by carefully crafted prompts. This encourages us to leverage LLMs for query extraction. Furthermore, the availability of open-source LLMs, such as the series of Llama (\cite{touvron2023llamaopenefficientfoundation,touvron2023llama2openfoundation}), ChatGLM (\cite{du2022glmgenerallanguagemodel,glm2024chatglmfamilylargelanguage}), Qwen (\cite{bai2023qwentechnicalreport,yang2024qwen2technicalreport}), allows us to conduct this study in a practical manner. When given a piece of toxic content, the useful parts for exploration may appear in various forms, such as phrases, fragments of sentences, emojis, or prompts. We expect that a finetuned LLM can directly identify and extract these components.

In this study, we propose an approach of LLM based query extraction aimed at toxic content exploration (QExplorer), which can compete with humans in exploration efficiency. It can be classified as a sequence-to-sequence generation method but differs in design from previous methods. The QExplorer involves a 2-stage training process: instruction Supervised Fine-Tuning (SFT) and preference alignment using Direct Preference Optimization (DPO). Feedback of the search system is used to construct preference data. The model training objective is aligned with the goal of making the extracted queries useful for toxic content exploration.

The contributions of this work are in the following aspects:

\begin{enumerate}
    \item We formulate the query extraction as a 2-stage LLM alignment problem innovatively. The preference alignment enables further adaptation of the query extraction model, achieving improved performance when using Qwen1.5-7B-Chat and ChatGLM3-6B as base models.
    \item We propose a finetuning framework named QExplorer for query extraction aimed at toxic content exploration. This framework seeks to extract more effective queries to look for toxic items in search system by integrating feedback from the search system into the training process.
    \item Offline experiments demonstrate that our method outperforms baseline models and humans in extracting effective queries. The Online deployment shows a significant increase in toxic items detection.
\end{enumerate}

\section{Related Works}
\subsection{Keyphrase Extraction}
Keyphrase extraction approaches can be broadly categorized into three main types. Early work in this area follows a pipeline approach, often employing a 2-step method (\cite{hasan-ng-2014-automatic,7805062,8852151,choi2023simckpsimplecontrastivelearning}). Initially, a set of candidate phrases is generated using heuristics, part-of-speech tags, n-grams, or other techniques. Subsequently, a ranking model is trained to order the candidates, with the top n phrases being selected as the extracted keyphrases. Sequence labeling, a technique widely used in Named Entity Recognition (NER) (\cite{lample2016neuralarchitecturesnamedentity,chiu2016namedentityrecognitionbidirectional}), has been adapted for keyphrase extraction over the past decade (\cite{10.1145/3308558.3313642,luan-etal-2017-scientific,zhang-etal-2016-keyphrase,Gu_2021}). In sequence labeling training, designed labels are assigned to each token to indicate whether it is part of a keyphrase. The model is then trained to predict the probability of each token should be part of the expected keyphrase. In recent years, sequence-to-sequence model have been employed to extract keyphrases (\cite{wu2024pretrainedlanguagemodelskeyphrase,chowdhury2022applyinggenericsequencetosequencemodel,9443960}). These models are typically built on pretrained language models such as Bert, Bart, T5. Our proposed approach can be considered as a sequence-to-sequence model, but it features a novel training design.

\subsection{LLM Alignment}
Since the launch of ChatGPT, the impact of LLM has been increasingly recognized. In addition to the extensive data used for training, two other valuable techniques for addressing various tasks are instruction SFT and preference alignment. A well pretrained LLM can be adapted to various downstream tasks via instruction SFT. These studies (\cite{wei2022finetunedlanguagemodelszeroshot,mishra2022crosstaskgeneralizationnaturallanguage}) have shown the zero-shot ability and generalization of instruction finetuned language models. This study (\cite{mishra2022crosstaskgeneralizationnaturallanguage}) observes that multi-task instruction finetuned language models exhibit cross-task generalization, with finetuned models showing improved performance on unseen tasks. This study (\cite{chung2022scalinginstructionfinetunedlanguagemodels}) demonstrates that using a series of designed instructions allows finetuned language models to achieve significant improvements. The study (\cite{ouyang2022traininglanguagemodelsfollow}) by OpenAI combines instruction finetuning and reinforcement learning from human feedback (RLHF) and shows an impacted improvement in the performance of LLM. Additionally, subsequent LLMs (\cite{bai2023qwentechnicalreport,touvron2023llama2openfoundation,peng2024largelanguagemodelbased}) have demonstrated the effectiveness of instruction finetuned LLM in query reformulation for search purposes.

The studies (\cite{ouyang2022traininglanguagemodelsfollow,touvron2023llama2openfoundation}) demonstrate that implementing RLHF after instruction SFT can greatly increase the win rate of LLM in generating tasks. Specific comparisons of the performance between LLMs with RLHF and without RLHF are presented in these studies. However, RLHF requires a reward model to be trained first, and the training of RLHF is relatively unstable. Alternative methods like DPO (\cite{rafailov2023directpreferenceoptimizationlanguage}) and odds ratio preference optimization (ORPO) (\cite{hong2024orpomonolithicpreferenceoptimization}) have been proposed. These methods do not require an explicit reward model; instead, the preference alignment is trained using pairs of preference data. This study (\cite{hou2024chatglmrlhfpracticesaligninglarge}) focuses on alignment of LLMs with human feedback, employing both DPO and PPO. DPO exhibits a promising performance in preference alignment, benefiting from a simpler design and requiring fewer resources during training.

Fully tuning the parameters of an LLM is expensive because it involves training all of the model's parameters. Methods such as LoRA (\cite{hu2021loralowrankadaptationlarge}), LoRA+(\cite{hayou2024loraefficientlowrank}) make the adaptation of LLM feasible with relatively low resources. During training, the parameters of the base model are frozen,  and trainable rank decomposition matrices are introduced to modify each layer of the transformer. These methods enable finetuning an LLM with a size of 7B or 14B using a single A100 GPU.

\section{Methodology}
\begin{figure}[htbp]
  \centering
  \includegraphics[width=\linewidth]{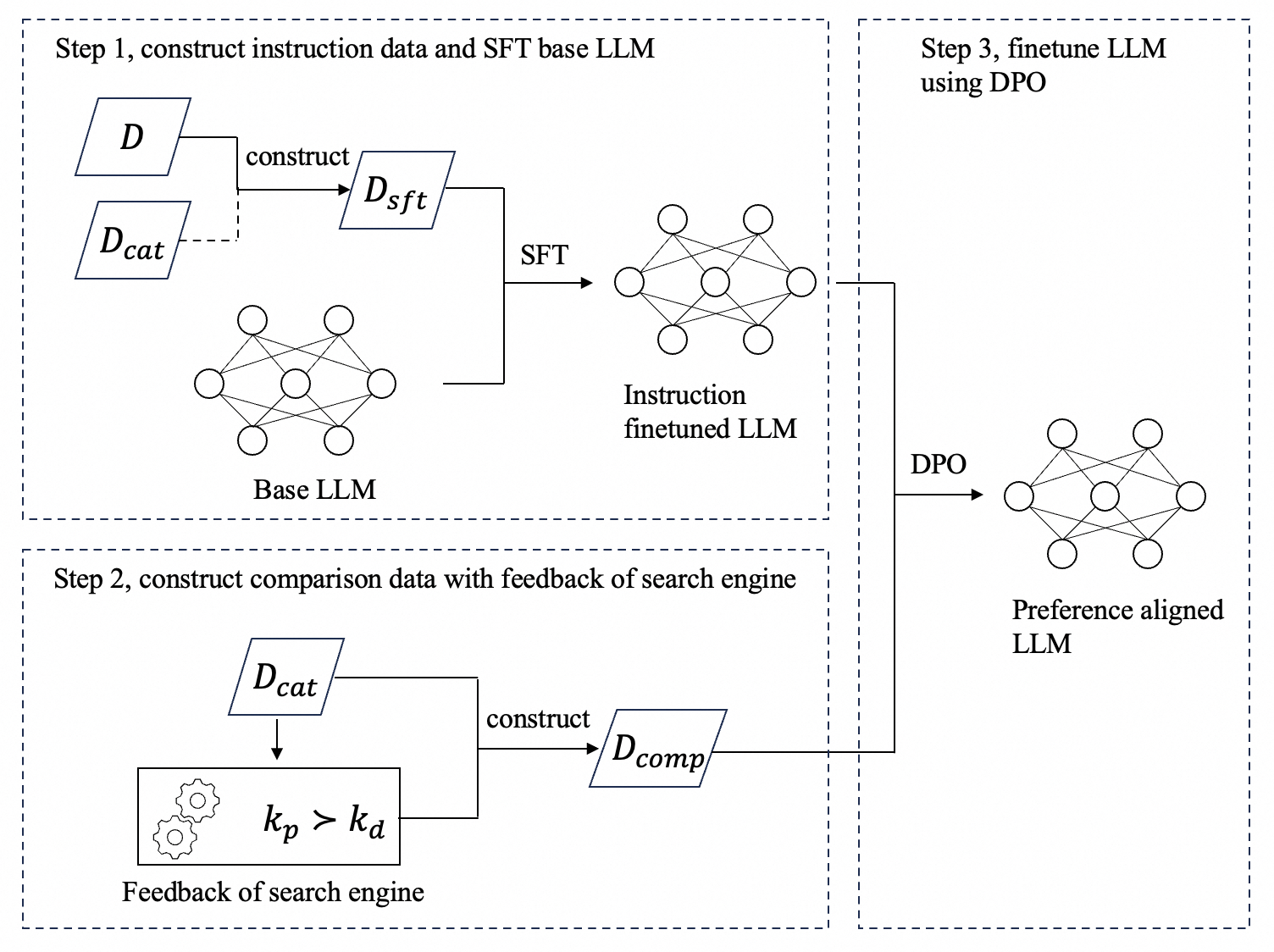}
  \caption{A diagram of QExplorer is presented. It contains three steps: (1) instruction SFT, (2) preference data construction, (3) preference alignment, which involves the well-finetuned LLM from step (1) using the preference data from step(2).}
  \Description{A diagram of our method.}
  \label{fig:method-diagram}
\end{figure}

Large Language Model based query extraction for toxic content exploration aims to get the key components within the given piece of content. These key components may include toxic expressions, suspicious prompts, or combinations of several rare co-occurrence phrases. The proposed approach intends to leverage the comprehension and reasoning abilities of LLMs. We propose a two-stage method, which includes instruction Supervised Fine-Tuning (SFT) and Direct Preference Optimization (DPO). In this study, we focus exclusively on the textual information of the reported toxic items. Figure \ref{fig:method-diagram} illustrates the diagram of our method.

\subsection{Instruction SFT}
Publicly available LLMs tend to refuse to engage with sensible content that involves offensive, sexual, violent or fraudulent languages. And they are not designed for extracting effective queries for further search. In the query extraction task, these models tend to list out all the possible phrases. On the other hand, we cannot use a third party API due to data security and cost concerns. Therefore, we create training datasets from the logs of our real production system and fine-tune LLMs to enhance their ability to extract effective queries from toxic content.

As shown in Figure \ref{fig:toxic_dec_system}, the auditors analyze the reported toxic content and mark out the improper parts during the risk understanding phase. Most of the annotated phrases are used to search for similar toxic content. The searched keywords, exposed items and detected toxic items are stored in the system log. Some typical examples are provided in Table \ref{tab:typical_toxic}. Not all annotated data is suitable for creating a training dataset. We select the annotated data with non-zero amount of exposed items or a length no more than 10. In this step, we include the annotated keywords regardless of whether they can lead to toxic items. This is because some keywords that do not yield toxic items merely indicate the absence of toxic items in the indices at the searching time, rather than being useless. We removed data where the annotated phrase length is less than 2 or greater than 10; these values are based on empirical observation. We denote the reported toxic content as $C$, the annotated keywords set as $K$, the number of exposed items caused by keyword $k_i$ in our search system as $hit(k_i)$, and the length of keyword $k_i$ as $len(k_i)$. The dataset used to for the instruction SFT can be formulated as follows:
\begin{equation}
  D = \left\{
    \left( c^i, k^i \right) \Bigg|_{\substack{i=1}}^N
    \begin{array}{ll}
    hit\left( k^i \mid S \right) > 0 \text{ or } len\left( k^i \right) \leq 10, \\
    c^i \sim p(x), \\
    len\left( k^i \right) > 1
    \end{array}
    \right\}
\end{equation}
where $p(x)$ denotes the toxic content distribution in Xianyu trading platform, $S$ denotes the Xianyu search engine.

In order to construct some long context data, based on the above dataset, we cluster the data into thousands of groups according to toxic content category and similarity. This data is used to construct the preference data and in the ablation study. For each group we sample up to 20 different items and concatenate the text information together using comma to form content $C$, and similarly, concatenate the qualified keywords using commas. This process can be formulated as follows:
\begin{equation}
D_{cat} = \left\{
    \left( \text{con}(c_s^i), \text{con}(k_s^i) \right) \bigg|_{\substack{i=1, s=1}}^{N_c, s=n}
    \begin{array}{ll}
    \left( c_s^i, k_s^i \right) \in G^i, \\
    D = \bigcup_{i=1}^{N_c} G^i 
    \end{array}
    \right\}
\end{equation}
where $G_i$ denoted the clustered content, $con(\cdot)$ represents the concatenation of the given text. This dataset $D_{cat}$ is further used to construct the preference alignment data.

We construct the final instruction SFT dataset using the above datasets and task specific prompts. An example is shown in Table \ref{tab:instruct_prompt}. As illustrated in this example, the content $C$ is fed into the input and the annotated keywords $K$ are fed into the output part.

\begin{table}
  \caption{Prompt example for keyword extraction for toxic content exploration. Most samples contain only one annotated query, as illustrated in this example. The prompt is designed to be compatible with multiple queries.}
  \label{tab:instruct_prompt}
  \begin{CJK}{UTF8}{gbsn}
  \begin{tabular}{p{\linewidth}}
    \toprule
    Instruction: 你是一个内容安全专家，需要从违规商品信息中提取出有助排查相似风险的关键词，关注和违规相关的独特信息。违规信息如下：\\
    Input: 学生……刷多门优惠力度大。软件/程序/网站开发教材建议某宝购入学起来更方便。\\
    Output: 提取的风险排查关键词集合如下，以英文逗号分隔：学生……刷 \\
    (Instruction: You are a content safety expert and need to extract key words from the improper item's information to help detect more similar risks, focusing on the unique information related to the violations. The improper information is as follows: \\
    
    Input: Students... cram for multiple courses with great deals. Suggest to buy textbooks on taoxx for more convenient learning of software/programs/websites development.\\
    
    Output: The collection of the extracted keywords for risk exploration is as follows, separated by commas: Students... cram.)\\
    \bottomrule
\end{tabular}
\end{CJK}
\end{table}

In the supervised finetuning process, the concatenation of the instruction and input in the prompt serves as the input $x$, whereas the output in the prompt serves as label $y$. The objective of our query extraction model is to maximize the conditional probability $p(y|x)$. The loss can be depicted as follows:
\begin{equation}
\mathcal{L}_{SFT}(\theta) = - \sum_{(x,y) \in D_{SFT}} \sum_{t=1}^{T} \log P(y_t | x, y_{<t}, \theta)
\label{eq:sft}
\end{equation}
where $D_{SFT}$ represents the dataset for instruction finetuning, $\theta$ denotes the model parameters, and $T$ denotes the length of the label sequence $y$.

We apply LoRA (Low-Rank Adaptation of Large Language Models) to finetune the LLMs.

\subsection{Preference Alignment}
This preference alignment aims to align LLM to the objective of identifying more similar toxic items. For a given reported toxic content, we assume the keywords leading to toxic items in a search engine are more effective than those that only retrieve normal items. To incorporate search engine feedback into training process simply, we employ the DPO (Direct Preference Optimization) algorithm. For each reported toxic content, it is necessary to construct two distinct sets of keywords, organized in a partial order. The DPO algorithm is designed to adapt the LLM parameters such that the likelihood of the preferred response is higher than that of the dispreferred response. Practically, there is usually one annotated query for a given reported toxic content. To obtain a pair of partially ordered keyword sets, we concatenate similar toxic content into a single, extended context paragraph, ensuring enough keyword candidates for ordering. Preference alignment data is constructed based on the dataset $D_{cat}$. The data construction process in this step could be formulated as follows:

\begin{equation}
\begin{array}{l}
D_{comp} = \\
\left\{
    \left( \text{con}(c_s^i), \text{con}(k_{ps}^i), \text{con}(k_{ds}^i) \right)
    \Bigg|_{\substack{i=1,s=1}}^{N_c, s=n}
    \begin{array}{l}
        \left( c_s^i, k_{ps}^i, k_{ds}^i \right) \in G^i, \\
        D = \bigcup_{i=1}^{N_c} G^i, \\
        p(k_{ps}^i \mid S) > 0.05, \\
        p(k_{ds}^i \mid S) \leq 0.05
    \end{array}
\right\}
\label{eq:preference}
\end{array}
\end{equation}

where $con(\cdot)$ denotes the concatenation of the given text, $k_{ps}^i$ represents the preferred keyword and $k_{ds}^i$ represents the dispreferred keyword. $p(\cdot)$ denotes the percentage of toxic items can be found among the logged exposed items by the given keyword on the search platform. $S$ denotes the search engine. Empirically, the threshold of $p(\cdot)$ is set to 0.05. In the subsequent ablation study, we analyze the model's performance using different thresholds. Table \ref{tab:preference_prompt} illustrates a prompt example at this stage. A complete example is shown in table \ref{tab:app_pref_data} in the Appendix.

\begin{table}
  \caption{Prompt example for preference alignment. The answer is a pair of constructed responses to the given question, with the first response being preferred over the second according to the condition specified in formula (\ref{eq:preference}).}
  \label{tab:preference_prompt}
  \begin{tabular}{p{\linewidth}}
    \toprule
    question: similar to the input of table \ref{tab:instruct_prompt}.\\
    
    system: similar to the instruction of table \ref{tab:instruct_prompt}. \\
    
    answer: [“The collection of the extracted keywords for risk exploration is as follows, separated by commas: Finacea”, \\
    
    “The collection of the extracted keywords for risk exploration is as follows, separated by commas: inacea 15\%ren” ] \\
    \bottomrule
\end{tabular}
\end{table}

In the preference alignment training, the concatenation of the content of the system and the question within the prompt constitutes the input $x$. The preferred answer is denoted as the label $y_p$, while the dispreferred answer is denoted as the label $y_d$. The initialized LLM acts as a reference model with fixed parameters $\theta_r$ during training. As derived in \cite{rafailov2023directpreferenceoptimizationlanguage}, the loss for DPO can be expressed as follows:
\begin{equation}
\begin{array}{l}
\mathcal{L}_{DPO}(\theta) = \\
    - \mathbb{E}_{(x,y_p,y_d) \in D_{comp}}
    \left[
    \log \sigma \left( \beta \log \frac{P(y_p | x, \theta)}{P (y_p | x, \theta_{r})}
    - \beta \log \frac{P (y_d | x, \theta)}{P (y_d | x, \theta_{r})} \right)
    \right] \\
\label{eq:dpo_loss}
\end{array}
\end{equation}
where $P(\cdot)$ represents the likelihood of the label with the given condition. The variable $\theta$ denotes the trainable LLM parameters, while $\theta_r$ denotes the fixed reference LLM parameters. $\beta$ is a hyper parameter that controlls the deviation from the reference policy as described in \cite{rafailov2023directpreferenceoptimizationlanguage}. To maintain the ability of generating outputs as an LLM, the loss function for preference alignment is formed as follows:
\begin{equation}
\mathcal{L}(\theta) = \gamma\mathcal{L}_{SFT}(\theta) + \mathcal{L}_{DPO}(\theta)
\label{eq:preference_loss}
\end{equation}
where $\mathcal{L}_{SFT}$ is as equation (\ref{eq:sft}), $\mathcal{L}_{SFT}$ is as equation (\ref{eq:dpo_loss}), $\gamma$ is a hyper parameter.

We also utilize LoRA to perform finetuning for DPO.

\subsection{Online Service}
\begin{figure}[htbp]
  \centering
  \includegraphics[width=\linewidth]{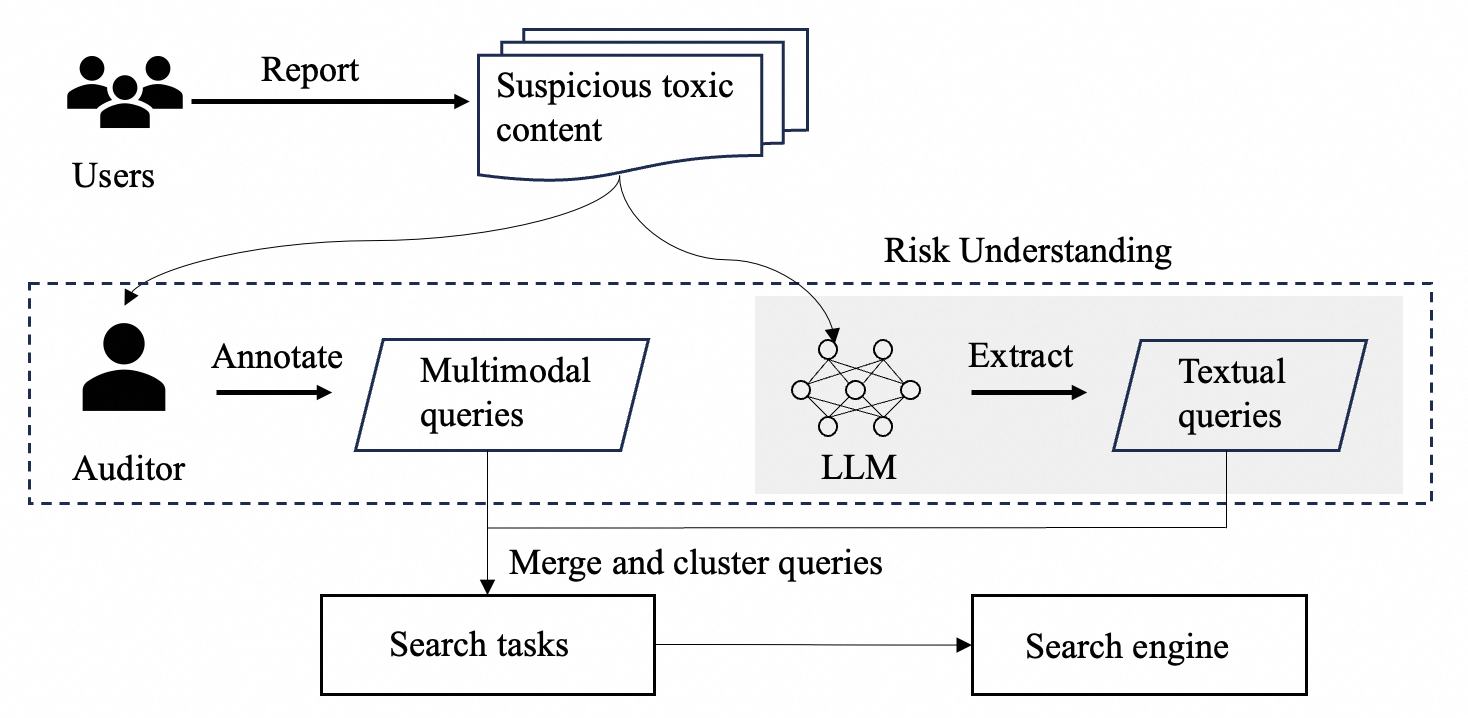}
  \caption{The diagram of the online implementation. The LLM in the shadowed part runs offline. Currently, this LLM can only process textual information. Therefore, the system still requires auditor to do risk understanding.}
  \Description{The diagram of online implementation.}
  \label{fig:online_implementation}
\end{figure}

The framework of our toxic content exploration system is illustrated in Figure \ref{fig:toxic_dec_system}. The aligned LLM is utilized to perform query extraction in the risk understanding phase, as shown in Figure \ref{fig:online_implementation}. This model complements human analysis. We deploy a data process pipeline on our Dataworks platform, which operates on a scheduled mode. Each day, the system receives a bunch of reported toxic items, and the textual information is fed into the finetuned LLM for inference. These reported toxic items are also analyzed by auditors. Both human annotated queries and LLM extracted queries are then used to create search tasks for further exploration. Each exploration task includes dozens of queries, with semantically similar queries are clustered into a single task. Whether a query originates from human annotation or LLM suggestion is logged so that we can assess the effectiveness of the proposed model. If an LLM suggested query matches a human annotated query, it is labeled as provided by a human. The online performance is evaluated in this implementation.

\section{Experiments}
\subsection{Datasets}
The data for training is processed from half year logs from our toxic content exploration system, covering the period of  [11.19. 2023, 05.19.2024]. We construct 25772 pairs of <toxic content, phrases> for dataset $D$, 2378 pairs long context samples for $D_{cat}$. Both $D$ and $D_{cat}$ are described in Section 3. The long context dataset $D_{cat}$ has an average length of 402 characters, while dataset $D$ has an average length of 154 characters. During training, the cutoff length of the input token sequence is set to 2048. For instruction SFT, we constructed 3 datasets using $D$, $D_{cat}$ and $D + D_{cat}$ respectively to study the effectiveness of different combination.

For preference alignment, using the method described in Section 3, we constructed about 600 preference samples.

Table \ref{tab:dataset_stat} shows the statistics of the datasets used in this study.

\begin{table}
  \caption{Statistics of datasets.}
  \label{tab:dataset_stat}
  \begin{tabular}{l|lll}
    \toprule
    Stage & Dataset & \# Samples \\
    \midrule
    \multirow{3}{*}{Instruction SFT} & $D$ & 25772 \\
    & $D_{cat}$ & 2378 \\
    & $D+D_{cat}$ & 27193 \\
    \midrule
    \multirow{2}{*}{Preference Alignment} & $D_{comp}$ & 619 \\
    & $D_{comp2}$ & 679 \\
    \midrule
    Test & $D_{reported\_20241020}$ & 402 \\
    \bottomrule
\end{tabular}
\end{table}

For the offline evaluation, we utilize the logged reported toxic items in our system after 05.19.2024. Using the same preprocessing as for training, we feed the text data into the finetuned LLM for inference. The offline test data presented in this study is derived from the reported toxic items of 10.20.2024.

\subsection{Offline Experiment Settings}
\subsubsection{Baselines} We compare QExplorer with typical methods used for keyword extraction, some LLMs and humans. The methods used for comparison are listed below.

\textbf{TF-IDF} is a simple method for selecting top N terms according to TF-IDF scores. Jieba segmentation is used in this study. N is set to equal the number of test samples.

\textbf{Bi-LSTM-CRF}(\cite{10.1145/3308558.3313642}) is a sequence labeling method. In our implementation, BIO tagging is used, where B and I tags are assigned to keywords, and words out of keywords are labeled to O. The Bi-LSTM-CRF model is trained on dataset $D$ for 500 epochs. Both the embedding size and hidden size are set to 256. The training data used in \cite{10.1145/3308558.3313642} consists more than 0.5 million documents, which is larger than the size of our SFT dataset $D$.

\textbf{BART}(\cite{chowdhury2022applyinggenericsequencetosequencemodel,lewis2019bartdenoisingsequencetosequencepretraining}) is a pretrained language model with encoder-decoder architecture for generation, translation, and comprehension tasks. It can be easily adapted to extract keyword and demonstrates competitive performance with state-of-the-art keyphrase generation on several benchmarks. In this study, bart-large is supervised finetuned to extract query with dataset $D$ for 5 epochs. Since BART cannot handle data more than 1024 tokens, the cutoff of input sequence is set to 1024.

\textbf{Base LLMs} A powerful LLM demonstrates impressive zero-shot capabilities across various tasks. However, in query extraction, a base LLM without SFT tends to list all the possible keywords. To ensure a fair comparison, we use a slightly modified prompt, as shown in Table \ref{tab:instruct_prompt}, for LLMs without SFT. In the prompt, we instruct the LLM to arrange the keywords in descending order of risk, then select the top 1 keyword for evaluation. Qwen2.5-72B-Instruct and GPT4 Turbo 128k are evaluated through local deployment, while GPT4 Turbo 128k is evaluated via the Azure OpenAI API. Other base LLMs used in this study, such as Qwen1.5-7B-Chat and ChatGLM3-6B, cannot follow the instructions well without SFT.

\subsubsection{Search Engine and Retrieval Setup} We do not build an isolated search engine for the offline evaluation, the online search engine is used. For a given set of reported toxic content, we create a series of retrieval tasks that includes all the queries extracted by LLMs or annotated by humans. The data retrieved from the search system is collected before the auditors start to explore. Once auditors begin their exploration, the search system becomes non-stationary because the toxic items are removed as soon as they are detected. The retrieval data collection takes about 1 to 2 hours, during which the search system can be considered approximately static. As the detection system depicted in Figure \ref{fig:toxic_dec_system} runs for 1 to 2 days after the retrieval data collection is completed, most of the toxic items in the system are labeled by auditors, we then use these toxic labels to calculate the offline metrics.

\subsection{Evaluation Tasks and Metrics}

\begin{table*}[h]
    \caption{Performance of different LLMs and humans. The best results are in bold, while the second-best results are underlined. The term"Number of hit@100" refers to the total amount of toxic items among the top 100 retrieved items. The term "\# Queries" refers to the number of extracted queries by each method with the same given content. "\# Effective Queries" refers to the number of queries that can retrieve toxic items.}
    \label{tab:main_result}
    \begin{tabular}{l|l|llll}
        \toprule
        Method & Number of hit@100 & \# Queries & \# Effective Queries & Query Hit Rate \\
        \toprule
        Human & 725 & 321 & 157 & 0.489 \\
        \midrule
        TF-IDF & 315 & 402 & 170 & 0.423 \\
        Bi-LSTM-CRF & 219 & 85 & 34 & 0.400 \\
        \midrule
        BART (SFT) & 799 & 347 & 134 & 0.386 \\
        \midrule
        Qwen2.5-72B-Instruct & 757 & 305 & 135 & 0.442 \\
        GPT4 Turbo 128k & 1082 & 247 & 130 & \ul{0.526} \\
        \midrule
        Qwen1.5-7B-Chat (SFT, LoRA) & \ul{1212} & 325 & 165 & 0.508 \\
        Qwen1.5-7B-Chat (QExplorer, LoRA) & \textbf{1321} & 296 & 166 & \textbf{0.561} \\
        \bottomrule
    \end{tabular}
\end{table*}

It is hard to set a gold standard for the query extraction for toxic content exploration. For a given reported toxic content, the most effective query for exploration depends on the items’ distribution of the search engine at a certain moment. Even an auditor may be unable to determine which query is optimal without conducting a search. For a given reported toxic item, an auditor can only assess whether the extracted query is acceptable based on their understanding. When an item is detected as toxic, it is removed from the search system. And in the online application, the human annotated queries have higher priority to search. Thus, we ave designed the following five criteria to evaluate the quality of the extracted queries.

Query hit rate, the percent of queries that could lead to the discovery of toxic items. This metric measures the model's ability to extract effective queries for toxic content exploration and is evaluated offline. 

Number of hit@100, the number of the toxic items found among the top 100 retrieved items in the search system using a given query. This metric is proportional to recall and measures the retrieval effectiveness of queries suggested by models or humans, and is evaluated offline. 

Query acceptance rate, depicts the chosen probability of the extracted queries by an auditor in online application, and is evaluated online. This metric is related to the offline query hit rate.

Toxic item detection increment, depicts the additional toxic items detected through the LLM suggested queries in an online application. This is evaluated online and relates to the offline number of hit@100. 

Hit query increment, depicts the additional queries that can retrieve toxic items compared to human, evaluated online.

\subsection{Implementation Details}
We utilize the open-source framework LLaMa-Factory (\cite{zheng2024llamafactoryunifiedefficientfinetuning}) for training. Qwen1.5-7b-chat (\cite{bai2023qwentechnicalreport}) and ChatGLM3-6B (\cite{glm2024chatglmfamilylargelanguage}) are used as the base models for finetuning. During the instruction SFT stage, the model is trained for 2 epochs with a learning rate of 3e-5. In the preference alignment stage, the finetuned model from the instruction SFT stage undergoes further training for 5 epochs with the same learning rate of 3e-5. The maximum length of the input token sequence is set to 2048. The LoRA method is used in the both stages, with the rank decomposition matrices employed for all the linear modules of a specific LLM. The optimizer is AdamW. The parameter $\beta$ in equation (\ref{eq:dpo_loss}) is set to 0.1. The parameter $\gamma$ in equation (\ref{eq:preference_loss}) is set to 1.0.

\subsection{Offline Effectiveness Analysis}
Table \ref{tab:main_result} presents the main results of human, baseline models and QExplorer in metrics of Number of hit@100 and Query Hit Rate. The proposed method Qwen1.5-7B-Chat (QExplorer, LoRA) outperforms all other models. Qwen1.5-7B-Chat (SFT, LoRA) is trained on dataset $D$. Qwen1.5-7B-Chat (QExplorer, LoRA) is further trained on the preference dataset $D_{comp}$ on base of Qwen1.5-7B-Chat (SFT, LoRA). The Query Hit Rate of Qwen1.5-7B-Chat (QExplorer, LoRA) is 0.561, which is higher than human labeled data by 14.7\%. It suggests that QExplorer can give more effective queries for toxic content exploration than humans. In metric Number of hit@100, it is observed that queries suggested by QExplorer can detect more toxic items than queries given by human. 

Large pretrained generative models outperforms TF-IDF and Bi-LSTM-CRF in query extraction. The TF-IDF and Bi-LSTM-CRF methods are less efficient than these generative models. TF-IDF tends to yield short queries, the efficiency of each query is not as high as other methods. Since the boundaries of queries in our situation are not strict, Bi-LSTM-CRF tends to assign more "O" tags to words within a sentence, resulting in fewer queries extracted compared to other methods. Bart is a powerful pretrained model; considering both Number of hit@100 and Query Hit Rate, BART (SFT) performs better than TF-IDF and Bi-LSTM-CRF. The results given by BART (SFT) is a strong baseline. The results given by GPT4 Turbo 128k and Qwen2.5-72B-Instruct are strong baselines too. Notably, GPT4 Turbo 128k surpasses human performance in this evaluation. However, GPT4 Turbo 128k suggests fewer queries than most of other methods, as it refuses to respond to questions involving offensive, sexual, violent content, etc. Nevertheless, GPT4 Turbo 128k does not surpass the finetuned Qwen1.5-7B-Chat. This observation indicates the necessity of aligning an LLM to a specific domain using customized data in practice. 


The comparison between Qwen1.5-7B-Chat (QExplorer, LoRA) and Qwen1.5-7B-Chat (SFT, LoRA) demonstrates that further search system preference alignment improves the Query Hit Rate by 10.4\% (0.508 v.s. 0.561), enabling the model to extract more effective queries. The objective of preference alignment is to generative queries that can retrieve toxic items, rather than merely producing queries that agree with human annotations, which is the goal of the instruction SFT.

\subsection{Ablation Study}

\begin{table*}[h]
    \caption{Ablation study of various base models, different SFT and preference alignment data.}
    \label{tab:ablation_study}
    \begin{tabular}{l|l|l|llll}
        \toprule
        Ablation Aspect & Method & \makecell{Number of\\ hit@100} & \# Queries & \makecell{\# Effective\\ Queries} & \makecell{Query\\ Hit Rate} \\
        \toprule
        \multirow{2}{*}{\makecell{Control Group}} & Qwen1.5-7B-Chat (SFT, LoRA) & 1212 & 325 & 165 & 0.508 \\
        & Qwen1.5-7B-Chat (QExplorer, LoRA) & 1321 & 296 & 166 & 0.561 \\
        \midrule
        \multirow{2}{*}{\makecell{Different\\ Base LLM}} & ChatGLM3-6B (SFT, LoRA) & 1183 & 321 & 151 & 0.470 \\
        & ChatGLM3-6B (QExplorer, LoRA) & 1232 & 304 & 154 & 0.507 \\
        \midrule
        \multirow{4}{*}{\makecell{Different\\ SFT Data}} & Qwen1.5-7B-Chat (SFT, LoRA, $D+D_{cat}$) & 1160 & 327 & 170 & 0.520 \\
        & Qwen1.5-7B-Chat (QExplorer, LoRA, , $D+D_{cat}$) & \textbf{1339} & 299 & 172 & \textbf{0.575} \\
        & ChatGLM3-6B (SFT, LoRA, $D+D_{cat}$) & 1309 & 326 & 162 & 0.500 \\
        & ChatGLM3-6B (QExplorer, LoRA, $D+D_{cat}$) & \ul{1332} & 296 & 170 & \ul{0.574} \\
        \midrule
        \multirow{2}{*}{\makecell{Different\\ Preference Data}} & Qwen1.5-7B-Chat (QExplorer, LoRA, $D_{comp2}$) & 1207 & 294 & 160 & 0.544 \\
        & ChatGLM3-6B (QExplorer, LoRA, $D_{comp2}$) & 1198 & 304 & 152 & 0.500 \\
        \bottomrule
    \end{tabular}
\end{table*}

In this section, we evaluate the performance of QExplorer by varying base LLMs and training data as shown in Table \ref{tab:ablation_study}.

\subsubsection{Different Base LLMs} ChatGLM3-6B has a different model architecture and uses different pretraining data compared to Qwen1.5-7B-Chat. The two models' size are comparable. As shown in Table \ref{tab:ablation_study}, the ChatGLM3-6B's performance is not as good as that of Qwen1.5-7B-Chat on this test dataset, but both LLMs demonstrate that further alignment with preference data results in a higher Query Hit Rate. 

\subsubsection{Different SFT Data} Table \ref{tab:ablation_study} shows that LLMs aligned within the QExplorer framework performs better than those aligned only through SFT. To determine whether improved performance is due to the long context samples not included in dataset $D$, we conducted experiments by varying the SFT data from $D$ to $D+D_{cat}$. Although Qwen1.5-7B-Chat (SFT, LoRA, $D$+$D_{cat}$) and ChatGLM3-6B (SFT, LoRA, $D$+$D_{cat}$) outperform their respective LLMs finetuned with dataset $D$, they still cannot surpass the performance of the corresponding LLMs aligned using the QExplorer framework (dataset $D$ is used in the instruction SFT stage). Further preference alignment improves the performance of both Qwen1.5-7B-Chat (SFT, LoRA, $D$+$D_{cat}$) and ChatGLM3-6B (SFT, LoRA, $D$+$D_{cat}$). We believe the QExplorer framework works in Qwen1.5-7B-Chat and ChatGLM3-6B when using this data recipe. We observe that the improvement provided by QExplorer on Qwen1.5-7B-Chat is more pronounced than on ChatGLM3-6B. This indicates that the ideal alignment recipe for various pretrained models may differ due to differences in model architecture, pretraining data, and training process. This comprehensive comparison was conducted after several months of online implementation. Although Qwen1.5-7B-Chat (QExplorer, LoRA, $D$+$D_{cat}$) has the best performance, our online service uses Qwen1.5-7B-Chat (QExplorer, LoRA) at the time of writing.

\subsubsection{Different Preference Data} In Section 3.2, the dataset $D_{comp}$ is constructed with a threshold of $p(\cdot)=0.05$. To assess the impact of the threshold for selecting preferred keywords, we constructed another dataset $D_{comp_2}$, where $p(\cdot)=0$. In dataset $D_{comp_2}$, a larger number of less efficient queries are selected as the preferred keywords. In Table \ref{tab:ablation_study}, the models in the "Different Preference Data" group use dataset $D$ for instruction SFT. We observe that dataset $D_{comp_2}$ can still improve the corresponding models' performance in Query Hit Rate, but not as effectively as dataset $D_{comp}$. In term of Number of hit@100, the dataset $D_{comp_2}$ fails to improve the performance of the model in the instruction SFT stage. 

\subsubsection{Preference Alignment Analysis} Table \ref{tab:ablation_study} shows that for the comparison between LLMs trained with SFT and the QExplorer framework, the LLM trained with our QExplorer framework performs better. This indicates that preference alignment can further improve the performance of LLMs in this query extraction task within the above experimental settings. However, not all DPO processes can enhance a model's performance; this study (\cite{xu2024dposuperiorppollm}) shows that DPO cannot surpass the SFT model performance on the code generation benchmark APPS (\cite{hendrycks2021measuringcodingchallengecompetence}). In this query extraction task, LLMs trained through SFT tend to generate a wider variety of queries compared to those trained using the QExplorer framework. This suggests that preference alignment reduces the diversity of the generated queries, which is expected, as queries differing only by auxiliary tokens are not helpful for exploration.

\subsection{Online Performance}

\begin{figure}[htbp]
  \centering
  \begin{subfigure}[t]{0.5\textwidth}
      \includegraphics[width=\linewidth]{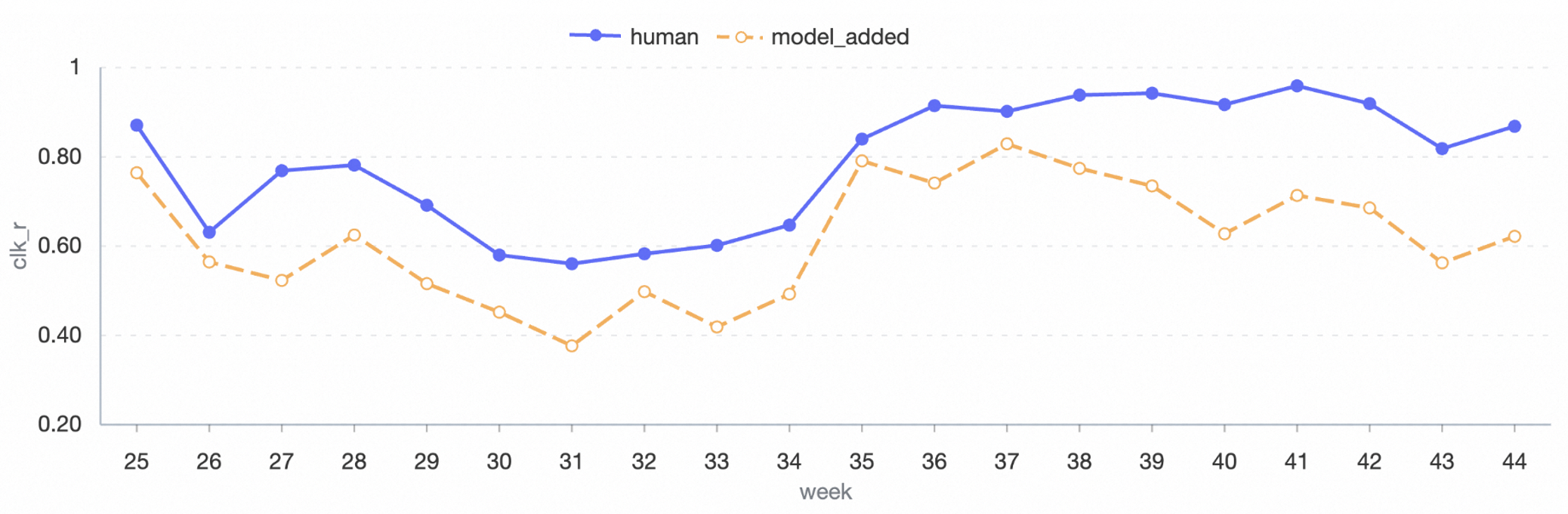}
      \caption{Query acceptance rate of our online service. The blue line with solid points indicates the result of human annotated queries. The orange line with hollow points indicates the result of the additional queries suggested by model.}
      \Description{Online query acceptance rate}
      \label{fig:online_performance_1}
  \end{subfigure}

  \begin{subfigure}[t]{0.5\textwidth}
      \includegraphics[width=\linewidth]{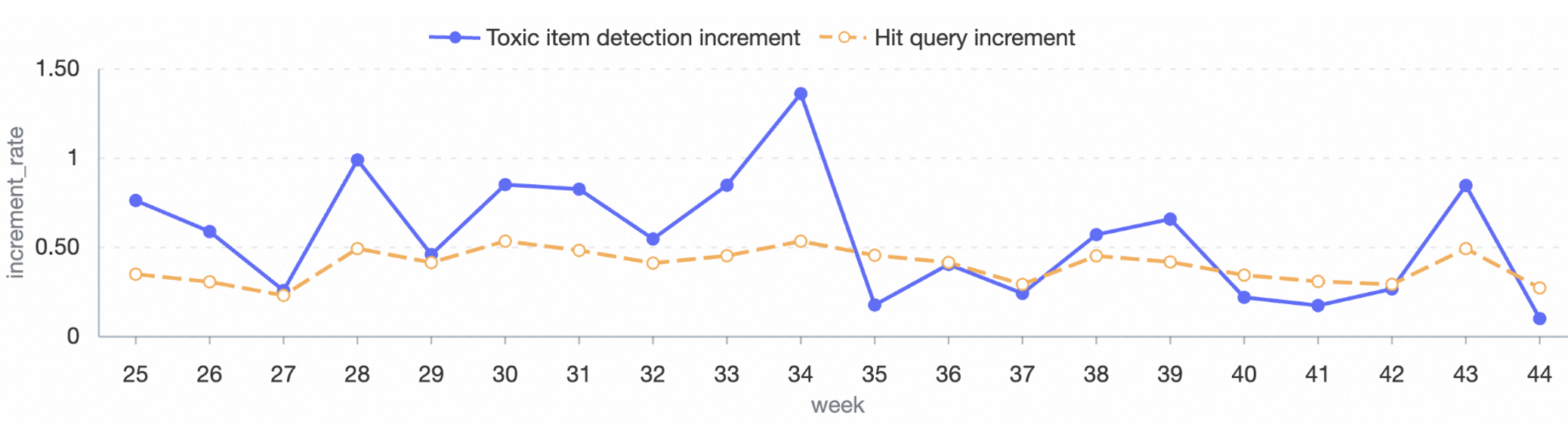}
      \caption{The vertical axis indicates the increment rate. The blue line with solid points indicates the toxic item detection increment given by model suggested queries. The orange line with hollow points indicates the hit query increment.}
      \Description{Online detection increment.}
      \label{fig:online_performance_2}
  \end{subfigure}
  \caption{Online performance over time.}
  \label{fig:online_perfrormance_graph}
\end{figure}

Following the offline validation of the effectiveness of the Qwen1.5-7B-Chat (QExplorer), we have been deploying this version of LLM in our online service since June 17th, 2024. The implementation details are in Section 3.3. The online performance demonstrated in this study is assessed using data collected over a period of 20 weeks (from 06.17.2024 to 11.02.2024). The results are shown in Table \ref{tab:online_effect} and Figure \ref{fig:online_perfrormance_graph}. The LLM suggested queries have an average acceptance rate of 61.6\%, which is competitive with that of queries given by humans, especially considering that queries overlapping between the LLM and humans are classified as human annotations. Approximately 20\% of the queries suggested by the LLM coincide with human annotations. The average toxic item detection increment is 59.9\%, indicating that the queries suggested by the LLM enable auditors to identify 59.9\% more toxic items. The average hit query increment is 39.9\%.

\begin{table}
  \caption{Online performance of the queries suggested by LLM and human.}
  \label{tab:online_effect}
  \begin{tabular}{l|llll}
    \toprule
    Method & \makecell{Query\\ acceptance\\ rate} & \makecell{Toxic item\\ detection\\ increment} & \makecell{Hit\\ query\\ increment} \\
    \midrule
    Human & 78.7\% & - & -\\
    \makecell{Qwen1.5-7B-Chat\\ (QExplorer)} & 61.6\% & 59.9\% & 39.9\%\\
    \bottomrule
\end{tabular}
\end{table}

\section{Conclusion and Discussion}
In the content safety area, one challenging issue is how to detect all the suspicious content with a limit amount of given toxic content. Xianyu is a second-hand trading platform, and each day a small number of toxic items are reported to our system. To maintain Xianyu as a healthy community, we need to address all these reported toxic items. Previously, the toxic content analysis was performed by auditors. In this study, we propose a method QExplorer to assist auditors to explore toxic items more efficiently. Firstly, we formulate this issue as a query extraction problem using LLM. Next, we construct two datasets from the logs of the online system with well-designed restrictions. To construct long context samples, we cluster the toxic items according to their category and similarity, then concatenate the similar content to form a reasonable long context. For preference alignment, we use the search engine as a reward model to select queries that are more effective. DPO is used to align the LLM with search system preferences. Both the instruction SFT and preference alignment are trained using LoRA. We introduce several metrics to evaluate the effectiveness of the proposed method, QExplorer. Results of offline experiments show that the queries suggested by Qwen1.5-7B-Chat (QExplorer) outperform the queries crafted by human. Online experiments reveal that Qwen1.5-7B-Chat (QExplorer) can suggest a substantial percentage of high quality queries and significantly enhance toxic item detection.

Our future work will focus on the following aspects: In this study, the data for the preference alignment is scarce. We can use beam search of LLM or results of different version of LLM, along with the feedback from search engine, to construct more data. Currently, we focus only on textual information for toxic content exploration. Visual information is also important, and multimodal finetuning of LLM can be considered in future.


\bibliographystyle{ACM-Reference-Format}
\bibliography{main}

\appendix

\section{Data Examples}
A sample from our preference alignment dataset is shown in table \ref{tab:app_pref_data}. 

\begin{table}
  \caption{An example of preference data. Our data is mainly in Chinese, the first line of the table is the original data. The English translation is provided in the second line. The content of question is constructed by combining the textual information from several semantically similar items. Thus there are enough keywords candidates for ordering, which ones are partially better than others. The first one in answer is supposed to be better than the second one for similar item exploration according to the feedback of our search system. The preferred query candidates are highlighted in red, while the dispreferred ones are highlighted in blue.}
  \label{tab:app_pref_data}
  \begin{CJK}{UTF8}{gbsn}
  \begin{tabular}{p{\linewidth}}
    \toprule
  \{
    "question": "违规信息如下：亚马逊猛鱼，鲳鱼，吃食嘎嘎猛。看上私聊发吃食视频 食人鱼 \textcolor{red}{红腹水虎},猛鱼嘎嘎猛，都是12CM以上，如果去网上买就要问老板能不能保证是三角\textcolor{blue}{锯齿}，不然就是假货感兴趣的话点我想要和我私聊吧,\textcolor{blue}{猛鱼水hu}小苗8公分左右。新手入门级P属性群体鱼 感兴趣的话点我想要和我私聊吧,\textcolor{red}{红腹}20多条，10公分左右，先到先得，不发货只能自提。感兴趣的话点我想要和我私聊吧,\textcolor{red}{红腹水虎}，15cm左右，青海民族大学自提，两条80元 感兴趣的话点我想要和我私聊吧,\textcolor{blue}{红腹食人鱼} 十厘米左右 食人鲳！感兴趣的话点我想要和我私聊吧", \\
    "answer": [ \\
      "提取的关键词集合如下，以英文逗号分隔：\textcolor{red}{红腹},\textcolor{red}{红腹水虎},\textcolor{red}{腹水虎}", \\
      "提取的关键词集合如下，以英文逗号分隔：\textcolor{blue}{猛鱼水hu},\textcolor{blue}{红腹食人鱼},\textcolor{blue}{锯齿}" \\
    ], \\
    "system": "你是一个内容安全专家，需要从违规商品信息中提取出有助排查相似风险的关键词，关注和违规相关的独特信息。" \\
  \} \\
  \midrule
  \{
    "question": "The improper information is as follows: Fierce Amazon Fish, Pomfret, aggressive eaters. If you're interested, send me a private message for feeding videos. Piranhas, \textcolor{red}{Red-bellied Water Tiger}, all are over 12CM. If buying online, make sure to ask the seller if they can guarantee it’s a triangular \textcolor{blue}{serration}; otherwise, it’s a counterfeit. If you’re interested, click me to chat privately. \textcolor{blue}{Fierce water taiger} Seedlings, approximately 8 cm. Beginner-level P attribute group fish. If you're interested, click to chat with me. 20+ \textcolor{red}{Red-bellied} Fish, Approximately 10 cm - First Come, First Served, Local Pickup Only. If you’re interested, click me to chat privately. \textcolor{red}{Red-Bellied Water Tiger}, Approximately 15cm, Self-Pickup at Qinghai Minzu University, 80 Yuan for Two. If you’re interested, click me to chat privately. \textcolor{blue}{Red-Bellied Piranha} - Approximately 10 cm - Piranha Fish! If you’re interested, click me to chat privately.", \\
    "answer": [ \\
      "The collection of the extracted keywords for risk exploration is as follows, separated by commas: \textcolor{red}{Red-bellied},\textcolor{red}{Red-bellied Water Tiger},\textcolor{red}{bellied Water Tiger}", \\
      "The collection of the extracted keywords for risk exploration is as follows, separated by commas: \textcolor{blue}{Fierce water taiger},\textcolor{blue}{Red-Bellied Piranha},\textcolor{blue}{serration}" \\
    ], \\
    "system": "You are a content safety expert and need to extract key words from the improper item’s information to help detect more similar risks, focusing on the unique information related to the violations." \\
  \} \\
  \bottomrule
\end{tabular}
\end{CJK}
\end{table}

\end{document}